\documentclass[conference]{IEEEtran}
\IEEEoverridecommandlockouts
\usepackage{cite}
\usepackage[utf8]{inputenc}
\usepackage{amsmath,amssymb,amsfonts}
\usepackage{algorithmic}
\usepackage{graphicx}
\usepackage{listings}
\usepackage{float}
\graphicspath{ {./gfx/} }
\usepackage{textcomp}
\usepackage{xcolor}
\usepackage{etoolbox}
\usepackage{subfigure}
\usepackage{hyperref}
\hypersetup{colorlinks,allcolors=black}
\definecolor{LightGray}{gray}{0.9}
\pdfoutput=1

\makeatletter
\newcommand{\linebreakand}{%
  \end{@IEEEauthorhalign}
  \hfill\mbox{}\par
  \mbox{}\hfill\begin{@IEEEauthorhalign}
}
\makeatother

\def\BibTeX{{\rm B\kern-.05em{\sc i\kern-.025em b}\kern-.08em
    T\kern-.1667em\lower.7ex\hbox{E}\kern-.125emX}}
    
\begin{document}

\title{Using Machine Learning at Scale in HPC Simulations with SmartSim:\\ {\LARGE An Application to Ocean Climate Modeling}}

%% Commented out for now to adhere to double-blind guidelines
\author{
 \IEEEauthorblockN{Sam Partee}
 \IEEEauthorblockA{\textit{\small{Hewlett Packard Enterprise}}\\
 Seattle, WA\\
 spartee@hpe.com\\
\small{ https://orcid.org/0000-0001-6005-5116}}
 \and
 \IEEEauthorblockN{Matthew Ellis}
 \IEEEauthorblockA{\textit{\small{Hewlett Packard Enterprise}}\\
 Seattle, WA\\
 matthew.ellis@hpe.com\\
 \small{https://orcid.org/0000-0002-5782-5447}}
 \and
 \IEEEauthorblockN{Alessandro Rigazzi}
 \IEEEauthorblockA{\textit{\small{Hewlett Packard Enterprise}}\\
 Switzerland\\
 alessandro.rigazzi@hpe.com\\
 \small{ https://orcid.org/0000-0003-2132-7726}} %Al did this. thanks Al
\linebreakand
\hspace{-.2cm} \IEEEauthorblockN{Scott Bachman}
 \IEEEauthorblockA{\textit{\hspace{-.2cm}\small{National Center for Atmospheric Research}}\\
 \hspace{-.2cm}Boulder, CO \\
 \hspace{-.2cm}bachman@ucar.edu\\
 \hspace{-.2cm}\small{https://orcid.org/0000-0002-6479-4300}}
 \and
 \hspace{.2cm}\IEEEauthorblockN{Gustavo Marques}
  \IEEEauthorblockA{\textit{\hspace{.2cm}\small{National Center for Atmospheric Research}}\\
\hspace{.2cm} Boulder, CO \\
\hspace{.2cm} gmarques@ucar.edu\\
 \hspace{.2cm}\small{https://orcid.org/0000-0001-7238-0290}}
 \and
 \IEEEauthorblockN{Andrew Shao}
 \IEEEauthorblockA{\textit{\small{University of Victoria}}\\
 Victoria, CA \\
 aeshao@uvic.ca\\
 \small{https://orcid.org/0000-0003-3658-512X}}
\linebreakand
\IEEEauthorblockN{Benjamin Robbins}
 \IEEEauthorblockA{\textit{\small{Hewlett Packard Enterprise}}\\
 Seattle, WA \\
 benjamin.robbins@hpe.com}
}

% DEADLINES
% February 28, 2020 – Submissions open
% March 31, 2020 – Abstracts submission deadline
%April 8, 2020 – Full paper deadline (No extensions; includes manuscript, AD, and optional AE appendix)
% May 9, 2020 – Reviews sent
% May 23, 2020 – Resubmission deadline (Includes manuscript, AD, and optional AE appendix)
% June 15, 2020 – Notifications sent
% July 15, 2020 – Major revision deadline (Includes manuscript, AD, and optional AE appendix)
% August 7, 2020 – Major revision notifications sent
% August 26, 2020 – Final paper deadline
%
\maketitle
\thispagestyle{plain}
\pagestyle{plain}

\begin{abstract}
We demonstrate the first climate-scale, numerical ocean simulations improved through distributed, online inference of Deep Neural Networks (DNN) using SmartSim. SmartSim is a library dedicated to enabling online analysis and Machine Learning (ML) for traditional HPC simulations. In this paper, we detail the SmartSim architecture and provide benchmarks including online inference with a shared ML model on heterogeneous HPC systems. We demonstrate the capability of SmartSim by using it to run a 12-member ensemble of global-scale, high-resolution ocean simulations, each spanning 19 compute nodes, all communicating with the same ML architecture at each simulation timestep. In total, 970 billion inferences are collectively served by running the ensemble for a total of 120 simulated years. Finally, we show our solution is stable over the full duration of the model integrations, and that the inclusion of machine learning has minimal impact on the simulation runtimes. \footnote{All of the source code, datasets, and models for experiments described in this work are open source and publicly available for download at \href{https://github.com/CrayLabs/NCAR\_ML\_EKE}{https://github.com/CrayLabs/NCAR\_ML\_EKE} }

\end{abstract}

% \begin{IEEEkeywords}
% component, formatting, style, styling, insert
% \end{IEEEkeywords}

\section{Introduction}

Advances in machine-learning (ML) algorithms have spurred research and development for combining data-driven approaches and traditional numerical  simulations to improve both efficiency and accuracy. The codebases of these numerical models are typically written in Fortran/C/C++ and run on high-performance computing platforms (HPC) via OpenMP and/or MPI parallelization. New software solutions are thus needed to connect these compiled language codebases to rapidly evolving ML and data analytics libraries, typically written in Python. Currently, the diversity of programming languages, dependence on file input/output (I/O), and large variance in compute resource requirements for scientific applications makes it difficult to perform  analysis, training, and inference with most ML and data analytics packages at the scale needed for HPC numerical simulations.

On its surface, the problem of being able to interface HPC applications with ML libraries is one of language interoperability and software interface design. However, for the full convergence of these two disparate paradigms, the true difficulty (and opportunity) in bridging these workloads needs to be reformulated in terms of data exchange. That is, how is data passed between a simulation and ML model at scale while making efficient use of heterogeneous computational resources? Current approaches to addressing this problem can be roughly broken down into two categories: offline (the ML and numerical components of a simulation do not exchange data directly) and online (the ML component is called while the simulation is running). Note that in this work, this definition of ''online" pertains to the process of inferring from a trained machine learning model, not continuously updating ML model parameters which is sometimes referred to as ''online learning".

To illustrate the differences between online and offline approaches, we review recent studies that couple ML and numerical models with a focus on computational fluid dynamics (CFD) and climate modeling domains due to the application presented in this work.

\subsection{Offline}

Offline ML surrogate modeling is the process by which a ML model is trained on data previously generated by a simulation. The surrogate is validated through the incorporation of surrogate inference data into a simulation through the filesystem. In this paradigm, a simple workflow would be to run the numerical simulation, store the output to disk, and train/validate a ML model on the stored result. 

The majority of efforts involving machine learning in climate modeling so far surround the use of surrogate models for turbulence parameterization or simulation component emulation (e.g. \cite{krasnopolsky_using_2013,brenowitz_prognostic_2018,han_moist_2020,beucler_achieving_2019}). In the climate modeling domain, an early example of this used an ensemble of neural networks (NN) to learn a stochastic convection parameterization \cite{krasnopolsky_using_2013}. A more recent study \cite{rasp_deep_2018} trained a deep neural network on one year's worth of data from the Super Parameterized Community Atmospheric Model version 3.0 to generate a ML-based parameterization of cumulus, deep convection. Both studies involved training a ML model on simulation data offline to create a surrogate model for a parameterization. One of the benefits reported by these studies is to reduce the time-to-solution as compared to reference simulations (used for training data) by replacing a numerically expensive portion of the simulation with an ML surrogate. However one of the major restrictions of the offline approach is that it is necessarily a nonconcurrent workflow, that is the ML and numerical simulations are completely decoupled, inhibiting the ability of each component to influence the other. In addition, for large scale simulations using file I/O storing diagnostics of the simulation can itself become a bottleneck, particularly for large-scale simulations.

\subsection{Online}

Online ML and numerical model applications indicate that inferences are performed as the numerical simulation is running. The authors stress the importance of this difference as offline and online performance are often significantly different \cite{brenowitz_machine_2020}.

Some ML libraries (e.g. Tensorflow and PyTorch) provide compiled language APIs so that ML can be ''hard-coded" into a simulation. For example, this approach has been used to enable ML solutions in a C++ based numerical CFD model, OpenFOAM \cite{openfoam}, by compiling in the C-based APIs for PyTorch \cite{geneva_quantifying_2019} and Tensorflow \cite{maulik_turbulent_2020}. However, these popular libraries do not include Fortran APIs and so this approach cannot be followed for the myriad of numerical models written in that language. While recent additions to the Fortran standard have formalized Fortran/C interoperability, the simulations themselves are often not written with such interoperability in mind, potentially requiring large refactors of simulation codebases and necessitating developers who are conversant in both C and Fortran. In addition, ML development generally evolves more quickly than than the numerical simulations due to the large-scale investment by industry and broader general interest. Developers and maintainers of numerical simulations must then divert their own relatively limited resources to maintain compatibility with ML libraries. 

Other approaches use the file-system as an intermediary between training a ML model and the inference process from within a simulation. For example, \cite{ogorman_using_2018} uses a random-forest to replace a parameterization of atmospheric convection in a climate model. The random forest model is trained offline and saved into NetCDF files which are loaded and called in the simulation by a custom Fortran module. Another approach, Fortran Keras Bridge (FKB) \cite{ott_fortran-keras_2020}, enables the usage of Keras-based ML models for online inference in Fortran. FKB builds on a NN library called Neural Fortran which implements a subset of modern ML methods in Fortran. To utilize FKB, Keras-based models must be trained, saved to file, and then loaded into Fortran simulations at runtime. These approaches are specific to Fortran and certain ML models/libraries, and involve similar requirements to embedding the TensorFlow or Pytorch C API into a simulation. Both approaches also lack the ability to utilize co-located or adjacent heterogeneous compute (CPU and GPU) capabilities that are becoming more prevalent in new HPC systems.

Another approach is rewrite the simulations (or portions thereof) in a ML-friendly language. For example, the CliMA project \cite{Schneider2017} relies heavily on the Julia programming language which by design generally has interoperability with Python. Given time and development resources, re-writing simulations in ML-friendly, performance oriented (e.g. not Python) languages is recognized as a potential avenue for the inclusion of ML in simulation at scale; however, many simulation codebases, developed over decades of research, are not as amenable to porting to a new language. 

In a study combining ML and the Finite Volume version 3 atmospheric generical circulation model \cite{mcgibbon_fv3gfs-wrapper_2021}, a Python wrapper was written to call the main timestepping routines. This approach has the advantage that the integration of the simulation can be controlled from Python. Additionally, they also provided interfaces to set and receive the state of the model via Python-Fortran interfaces  allowing for the direct diagnosis of the simulation state during the course of the simulation (e.g. online analysis). Writing these interfaces however required significant and specialized technical expertise. An additional complication of using a Python driver to interface with the underlying Fortran model is that Python's library management system scales poorly when hundreds or thousands of clients are importing libraries stored on a network-mounted filesystem \cite{feng2016} (containers can partly mitigate this problem but add complexity to the running of the application and may not supported on all HPC platforms).

\subsection{Outline}
In this paper, we describe a new software framework SmartSim that enables the convergence of numerical simulations and ML workloads on heterogeneous architectures. Synthetic scaling studies are performed to demonstrate the performance of the system. We then embed SmartSim into an ocean climate model that uses online inference to augment an existing turbulence parameterization. The computational performance and stability of this approach is evaluated by integrating a 12-member ensemble of global ocean simulations for 10 simulated years. Finally, we conclude by discussing the broader applicability of SmartSim beyond online ML inference.

\section{SmartSim Architecture}\label{section:smartsim_architecture}

\begin{figure*}[ht]
\centering
\includegraphics[width=0.9\textwidth]{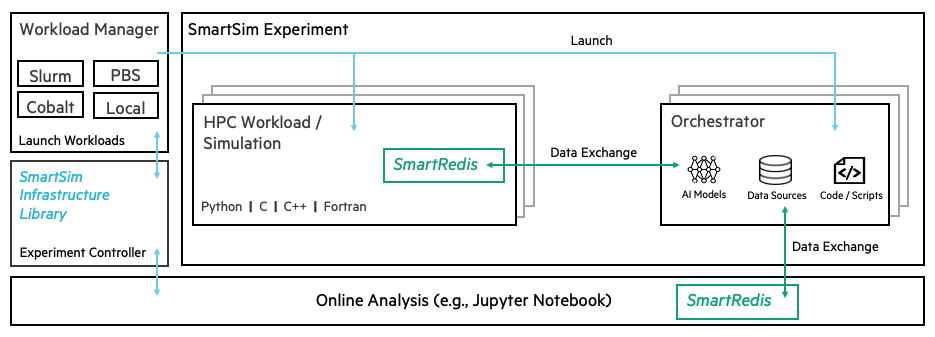}
\caption{The architecture for SmartSim is provided for a given use case. In this instance, the 
Infrastructure Library (IL) is being used to launch the Orchestrator alongside a simulation embedded with the SmartRedis clients. In addition to communication with the simulation, the Orchestrator is also sending data to an analysis environment for online analysis, visualization and/or training of a machine learning model. }
\label{fig:smartsim_arc}
\end{figure*}

SmartSim enables simulations in Fortran, C, C++ and Python to execute ML models hosted within in-memory storage (DRAM), facilitating online inference at simulation runtime. In addition, SmartSim's API orchestrates the movement of data between simulation and ML components with simple put/get semantics. SmartSim can host ML models on CPU-only or GPU-enabled compute nodes adjacent to or co-located with the simulation. SmartSim is portable and can be run on laptops and scales to thousands of processors as shown in this work.

SmartSim is comprised of two libraries: a lightweight client library, SmartRedis, that is compiled into end-users simulation, and an Infrastructure Library (IL) that facilitates the execution of simulations and deployment of in-memory storage. 

\subsection{SmartSim Infrastructure Library}

The Infrastructure Library (IL) provides a Python API to automate the process of deploying the components necessary to execute ML-augmented simulations, perform online analysis, or otherwise utilize the full feature set of SmartSim. The IL is designed for interactive computation in Jupyter environments, but can be executed as a Python application or script. The main motivation for the creation of the IL was to reduce the complexity of launching SmartSim-enabled workloads on HPC systems. 

The IL loosely integrates with existing system workload managers to deploy user-specified applications in conjunction with the aforementioned computational, in-memory storage. Slurm, PBSPro, and Cobalt are all supported by the IL. Local launching is also supported for single-node, workstation, or laptop workloads. If a system is not supported by the IL, or a workload requires specific launch methods, the infrastructure deployed by the IL can be initialized manually.

The application presented in section \ref{application-results} utilized the IL to deploy machine
learning infrastructure as well as the SmartRedis-enabled ocean models. The entire workflow was conducted from a Jupyter lab environment using the IL python interface.

\subsubsection{Models and Ensembles}

The \texttt{Experiment} object is the primary user interface of the IL. Experiments are used to create references to HPC applications referred to as a \texttt{Model}. Model instances contain all parameters for executing a given application on a system, such as compute resource parameters, model parameters, and input files. Model instances are provided compute resource parameters in the form of \texttt{RunSettings} objects. The IL provides multiple types of settings that parameterize execution on systems with workload managers. For example, \texttt{SrunSettings} are used on systems that utilize the Slurm scheduler. More general launch binaries, such as \texttt{MpirunSettings} for OpenMPI, are also supported by the IL. 

In addition to models, users can create \texttt{Ensemble} objects. Ensemble instances are collections of \texttt{Model} instances that are treated as a single workload. Ensembles create models upon initialization by generating sets of provided simulation parameters (ex. number of timesteps) and writing those parameters into specified input files. Ensembles can instead create replicas of models or be constructed manually through the addition of user-created models.

Ensembles are provided the same \texttt{RunSettings} instances which dictate how each ensemble member should be executed; however, ensembles can also accept \texttt{BatchSettings} objects which signify that the entire ensemble should execute as one batch. For example, \texttt{QsubBatchSettings} parameterize the execution of batch workloads on systems that utilize the PBS scheduler.

Once created, Model and Ensemble instances are used as references to workloads that can be started, monitored, stopped, and restarted through the experiment interface. The experiment allows users to launch workloads asynchronously such that after applications are successfully launched users can continue to execute code. This launching paradigm is crucial for interactive environments like Jupyter Notebooks for which SmartSim was designed. 

\subsubsection{Orchestrator}

The primary role of the IL is to launch in-memory storage alongside HPC applications. This storage is referred to as the Orchestrator in SmartSim and is built on Redis\cite{redis} and the Redis module system.

Once an Orchestrator instance is created, the experiment interface launches underlying Redis shards across compute nodes of a system and connects the nodes together in a shared-nothing cluster that users and applications can address as a single storage device. Orchestrators, like ensembles, are launched in preexisting or interactive allocations or as a batch job.

Combined with the SmartRedis clients, the Orchestrator is capable of hosting and executing ML models written in Python on CPU or GPU. Notably, despite being written in Python, all models are just-in-time compiled and executed in a C runtime. The Orchestrator supports models written with TensorFlow, Keras, Pytorch, TensorFlow-Lite, or models saved in an ONNX format (sci-kit learn). A Redis module, RedisAI, provides the aforementioned ML runtimes, creating a library agnostic layer between SmartRedis enabled simulations and ML libraries. Because of this, users of SmartSim can easily switch between ML frameworks and implementations without needing to make any changes to the simulation code.

Redis was chosen for the Orchestrator because it resides in-memory, can be distributed on-node as well as across nodes, and provides low-latency access to many clients written in different languages. With the SmartRedis clients, users can seamlessly connect Fortran, C, C++ and Python applications at scale without passing through the filesystem. A hashing algorithm, CRC16, is used to ensure that data is evenly distributed amongst all database nodes. Notably, a user is not required to know where (which database node) data or Datasets (see Dataset API below) are stored as the SmartRedis clients will infer their location for the user. 

The importance of the IL is two fold. First, the deployment capabilities greatly reduce the complexity and overhead of creating computational setups for the development of ML-augmented simulations. By enabling the programmatic parameterization and execution of applications, users can quickly generate workloads that explore large parameter spaces, perform uncertainty quantification, generate training datasets, conduct scaling tests, and more. Second, the infrastructure deployed by the IL, namely the Orchestrator, facilitates the connection of applications through the SmartRedis clients for many types of online workloads: inference, training, analysis, visualization, etc.

\subsection{SmartRedis} \label{section:smartredis}
    
   \subsubsection{Tensors}
   The SmartRedis clients use a n-dimensional tensor data structure for transferring data, storing data, evaluating scripts, and evaluating models. However, to minimize the code changes in applications, the SmartRedis tensor data structure is opaque to the user and native n-dimensional arrays in the host language (C, C++, Fortran, Python) are used in the user-facing API functions.  For example, in Python, the SmartRedis client works directly with NumPy arrays. For C and C++, both nested and contiguous memory arrays are supported. Only contiguous arrays are supported in Fortran, but with care taken to preserve the row-major convention. The reader is encouraged to consult the SmartRedis documentation for a detailed description of supported data types and API functions.
   
   \subsubsection{Datasets}
   In many scientific applications, multiple n-dimensional tensors are naturally grouped together as they have some contextual relationship.  Additionally, there is often metadata about the tensors (e.g. dimension names) or the simulation from which they come (e.g. time step information) that should be stored alongside the tensors.  The SmartRedis DataSet API allows users to group n-dimensional tensors and metadata into a single data structure that can be accessed or manipulated in the Orchestrator with a single key.  Specifically, users need not know where the tensors and metadata within the DataSet object are stored once they have been sent to the Orchestrator. Users only need to know the name given to the dataset when constructing the DataSet object.

  \subsubsection{Data Processing}
   The ability to perform online data processing is essential to enabling online inference. Most machine learning algorithms require some preprocessing of input data. Sometimes this processing is as simple as data normalization, but often more computationally expensive data processing is needed.
   
   SmartRedis provides an API for storing, retrieving, and executing TorchScript programs inside of the Orchestrator database.  The scripts are JIT-traced Python programs that can operate on any tensor data stored in the Orchestrator and execute on CPU or GPU. After the TorchScript execution, the output tensors of the script are stored in the Orchestrator and are accessible with a user-specified name. Such calls to the SmartRedis API can be chained together to processing and inference pipelines.

  \subsubsection{Model Inference}

   The SmartRedis clients support the remote execution of Pytorch, TensorFlow, Keras, TensorFlow-Lite, and ONNX models that are stored in the Orchestrator. With this capability, embedded SmartRedis clients can augment simulations with machine learning models stored in the in-memory database.
    
    SmartRedis clients support storing, retrieving and executing ML models with the aforementioned ML frameworks. When a call to \texttt{client.set\textunderscore model()} is performed, a copy of the model is distributed to every node of the database to leverage all available hardware. When performing the remote execution of a model through a SmartRedis client, the model chosen for execution is the model co-located with some or all of the model input data.  In the case where all of the input data or output for a model execution is not on the same node of the database, the SmartRedis client will move temporary copies of the input or output data to the node.  The movement of data between nodes for model inference is completely opaque to the user and is handled internally by the SmartRedis client.
    
    ML models stored in the Orchestrator can be executed on CPU or GPU. The output of models
    is stored within the Orchestrator until requested by the user. In this way, scripts and models can be executed sequentially to form computational pipelines for distributed processing and inference.

\section{Scaling SmartSim Applications} \label{section:benchmarks}

A synthetic scaling simulation was run on a Cray XC50 to quantify Orchestrator inference performance.  In this scaling analysis, a set of synthetic simulations was run with a fixed number of Orchestrator (database) nodes and varying number of SmartRedis clients and another set of synthetic simulations was run with varying number of database nodes and a fixed number of SmartRedis clients.  During the synthetic simulation, each client connected to the Orchestrator repeatedly sets a tensor, runs a PyTorch script to modify the tensor, runs a PyTorch model, and retrieves the output tensor of the PyTorch model.  Through this repeated set of SmartRedis client API calls, the ability of the SmartSim infrastructure to handle uncoordinated tensor, model, and script requests on a busy network can be assessed.

\subsection{Hardware Configuration}
The number of database nodes in the scaling experiment was varied from 4 to 16.  Each database node was equipped with an Nvidia P100 GPU, 64 GB DDR4-2400 memory, and 18-core 2.3 GHz Intel Broadwell processors.  The number of clients simultaneously connected to the database varied from 960 to 7,680 using computational nodes containing 48-core 2.1GHz Intel SkyLake CPUs with 192GB DDR4-2666 memory. The Cray XC50, used for this scaling study, utilizes the Aries interconnect. 

RedisAI, the Redis module that provides the ML runtimes to the Orchestrator, contains several methods for tuning performance for given workloads such as the number of I/O threads, GPU and CPU worker threads, and background threads. In this study, 4 threads per GPU (one P100  per compute node), one I/O thread, and one background thread were unbound and free to schedule on the 18 cores of the Broadwell CPU.

\subsection{Software Configuration}

\begin{figure}[!htbp]
    \centering
    \includegraphics[width=\columnwidth]{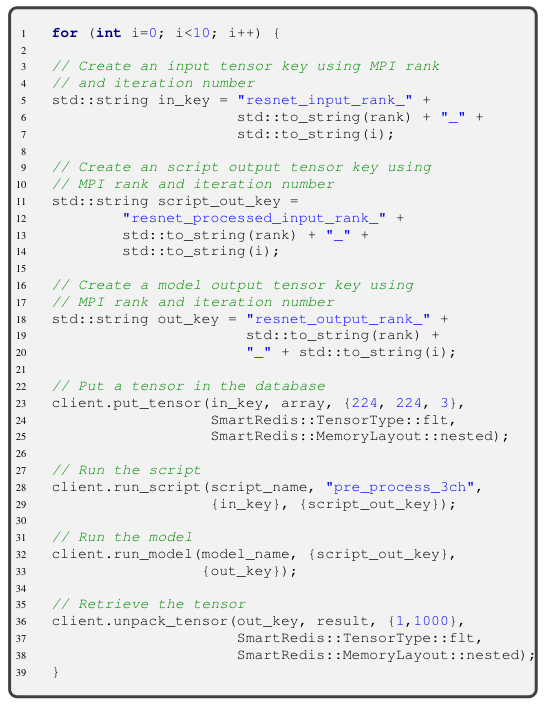}
    \caption{Excerpt of synthetic scaling study main loop}
    \label{listing:cpp_scaling_code}
\end{figure}

The synthetic scaling simulation is a C++ application that utilizes the SmartRedis client API to set tensors, execute models, execute scripts, and retrieve tensors.  The code snippet in Listing \ref{listing:cpp_scaling_code} shows the primary loop in the synthetic scaling simulation that contains the SmartRedis API calls.  Note that during the scaling study, each SmartRedis API call and the outer loop were enclosed by timing calculations, but these have been removed from the excerpt to improve readability.  Also, note that the code excerpt shows that a total of ten iterations of the SmartRedis API calls is performed by each MPI rank, which is consistent with the results that will be shown in Section \ref{section:scaling_results}.

The synthetic simulation code excerpt is executed by all of the MPI ranks in the scaling study.  Note that the model and script referenced in the code excerpt API calls is set by the SmartSim Python script through the SmartRedis Python client API.  In this way, the model and script are set in the database before the C++ synthetic simulation is executed.  The only optional parameter specified when setting the model with \texttt{client.set\textunderscore model()} is a batch size of 10,000.  By setting a large value for the batch size, the RedisAI module will group together tensors that arrive close together into a single model execution.  The ResNet-50 convolutional neural network was used in all of the synthetic scaling simulations.

The synthetic scaling simulation (C++ application) is executed by a SmartSim Python script that sets the total number of SmartRedis clients connecting to the database and the number of databases in each execution of the C++ application. For every permutaiton reported here, a fresh database cluster is launched to maintain uniformity across all executions.

\subsection{Results}
\label{section:scaling_results}
\begin{figure*}[htbp!]
    \vspace{-0.5cm}
    \begin{subfigure}
        \centering
        \includegraphics[width=\columnwidth]{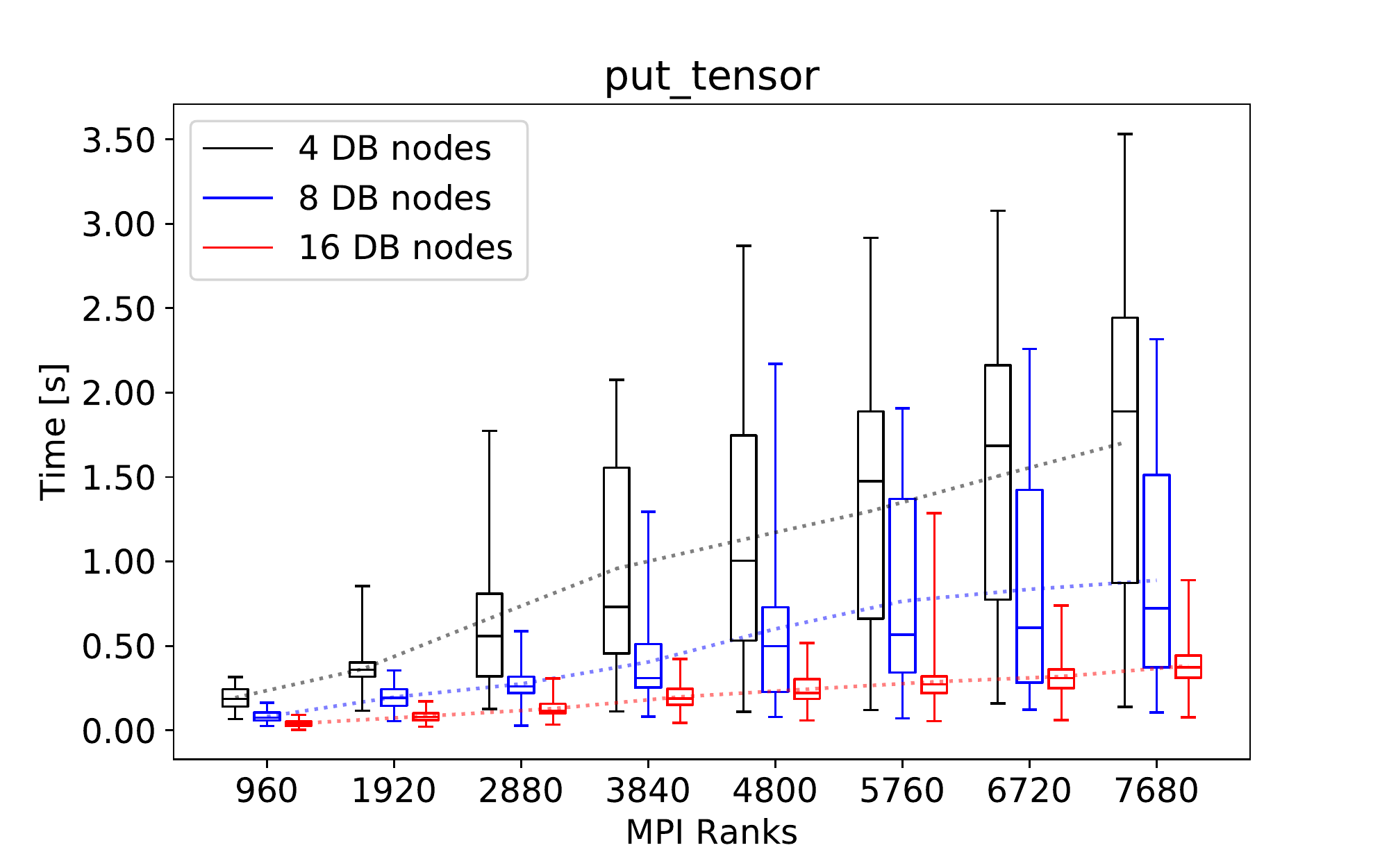}
    \end{subfigure}
    \begin{subfigure}
        \centering
        \includegraphics[width=\columnwidth]{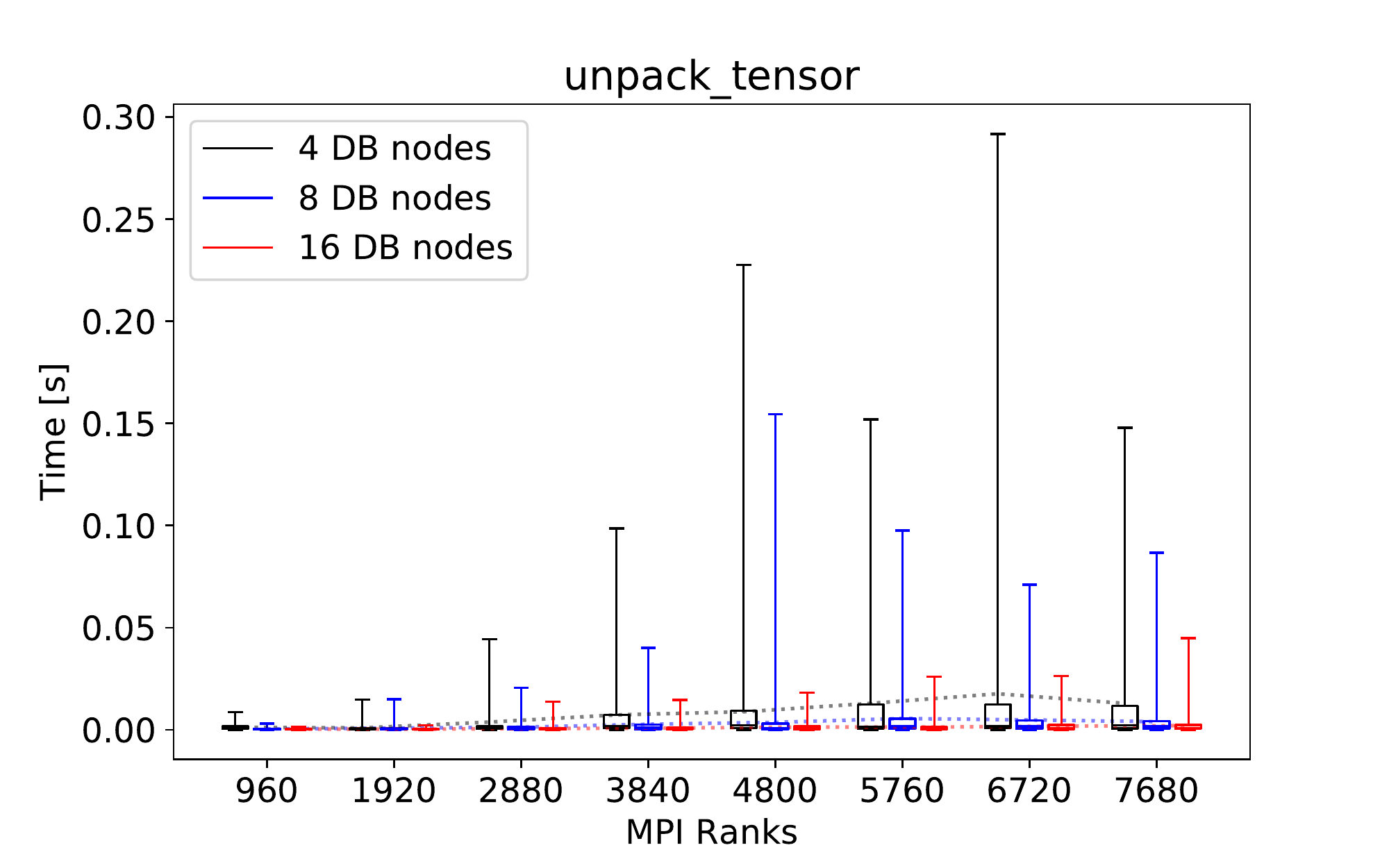}
    \end{subfigure}
        
     \vspace{-0.5cm}
     \begin{subfigure}
         \centering
         \includegraphics[width=\columnwidth]{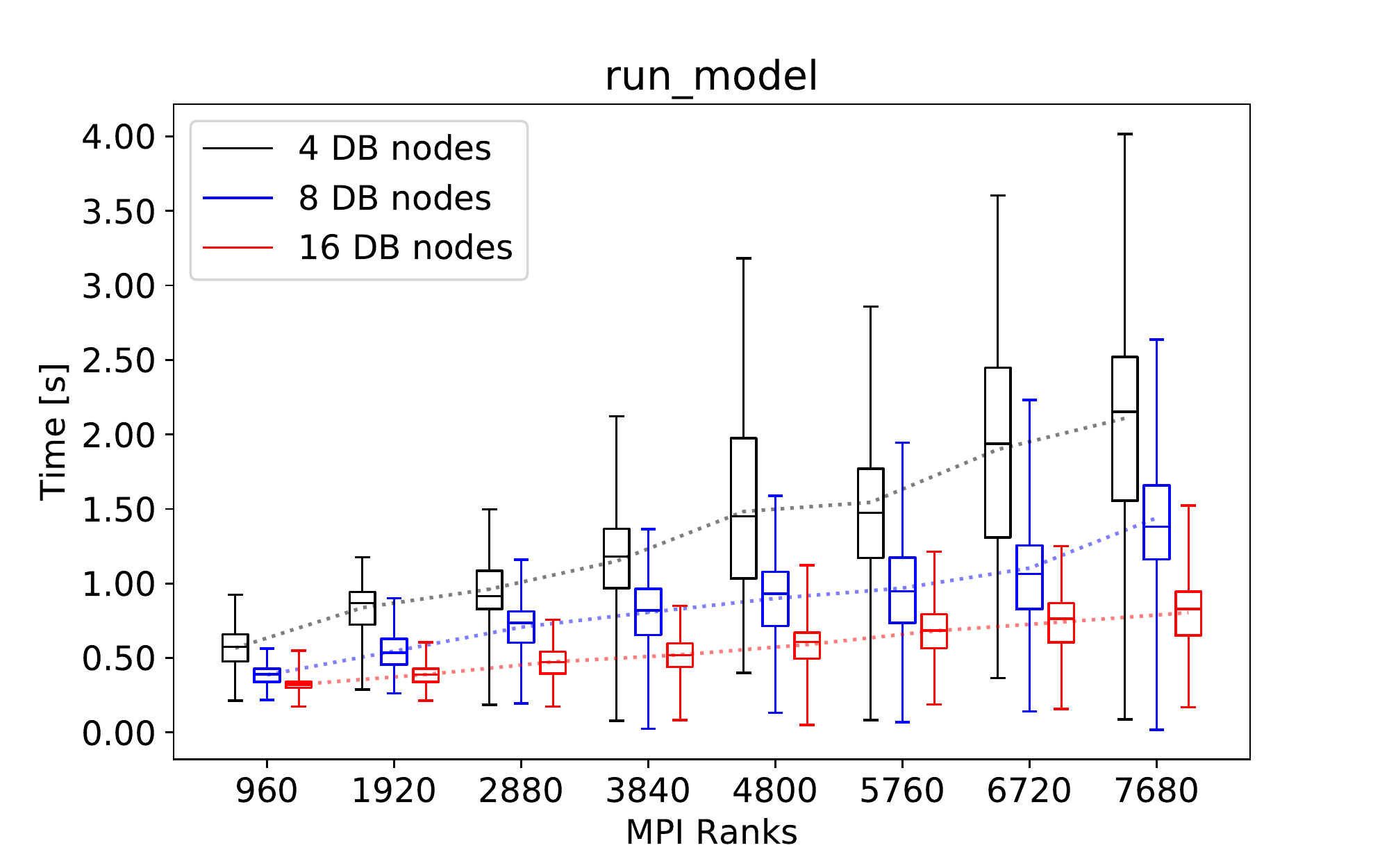}
     \end{subfigure}
     \begin{subfigure}
         \centering
         \includegraphics[width=\columnwidth]{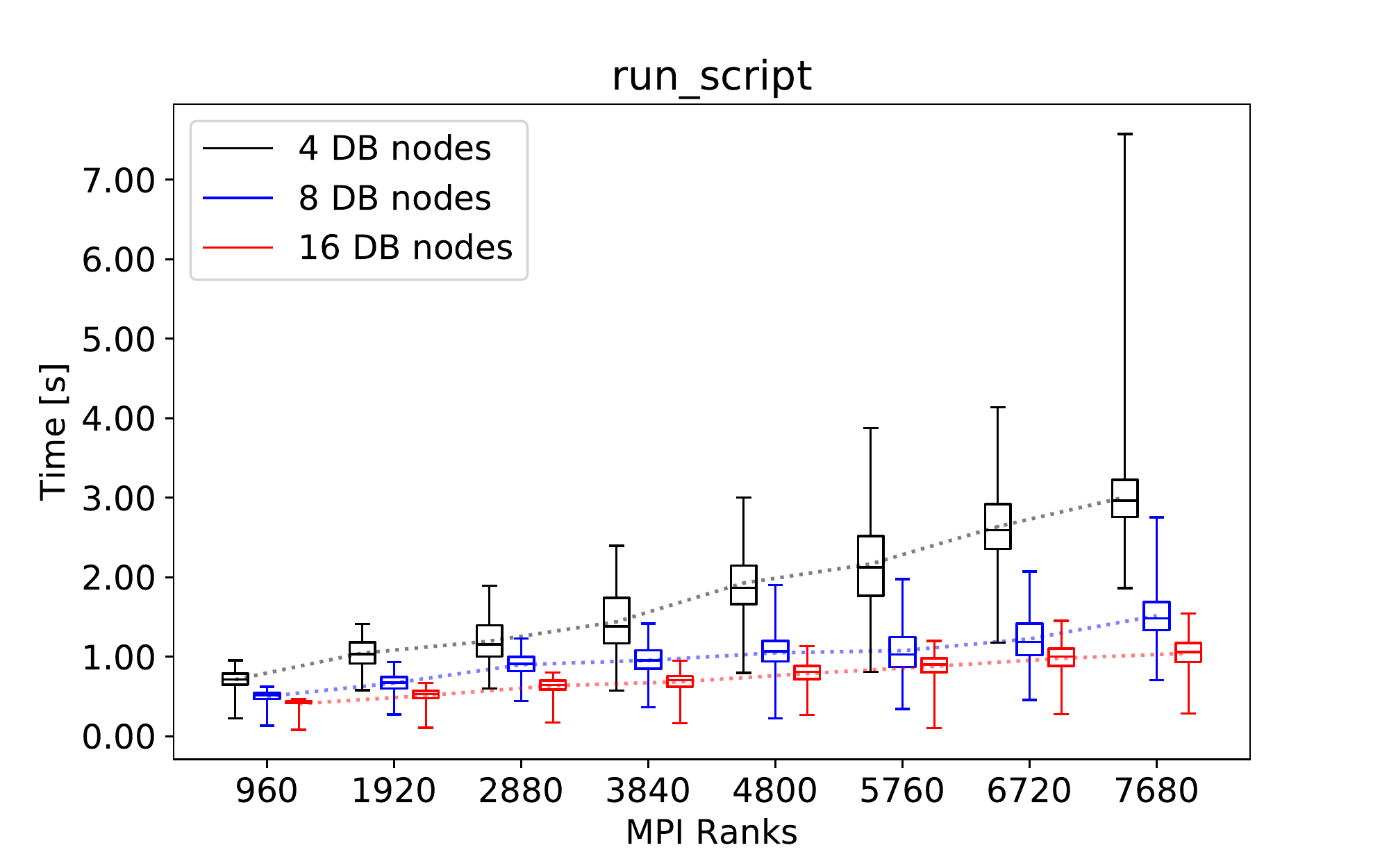}
     \end{subfigure}
        \caption{Execution time of SmartRedis API calls using Redis clusters of 4, 8, and 16 DB nodes, with varying client connection count. Box plots show median, first and third quartiles of timing distributions; box plot whiskers extend from minimum to maximum recorded timings. Dashed lines connect mean values. All code and instructions to reproduce these scaling
        experiments are open source and can be found at \href{https://github.com/CrayLabs/SmartSim-Scaling}{https://github.com/CrayLabs/SmartSim-Scaling}}
    \label{fig:timing_all_in_one}
\end{figure*}

Figure~\ref{fig:timing_all_in_one} show the scaling behavior of SmartRedis API calls for select combinations of database node counts and SmartRedis client counts.  Recall that in each run of the synthetic simulation there are between 960 and 7680 MPI ranks, and each MPI rank has a SmartRedis client connection to the database.  Additionally, each MPI rank executes ten iterations of SmartRedis API calls.  As a result there is a distribution of runtime across and within each MPI rank, with the aggregate behavior across all ranks and iterations are shown in the same plots.

In Figure~\ref{fig:timing_all_in_one}, the mean runtimes for varying client counts with fixed database node counts are connected with dashed lines.  At each data point, box plots show the median, first and third quartiles of timing distributions.  Additionally, box plot whiskers extend from minimum to maximum recorded timings. For the 3 most time-consuming API calls, \texttt{put\_tensor}, \texttt{run\_model}, and \texttt{run\_script}, the mean runtime of SmartRedis API calls scales linearly as the number of SmartRedis clients is increased with a fixed database node count. Adding additional database nodes reduces both the mean and maximum times. The reduction in maximum time of a rank is especially important for applications where \texttt{run\_model} would be a blocking part of the algorithm. % Moreover, the Table \ref{tab:efficiency} shows that performance significantly improves as the database count increases, but this improvement saturates at 16 database nodes.  It is expected that the performance values will improve significantly as optimizations like command pipelining and directed acyclic graphs (DAG) are introduced into SmartRedis.

\section{Application: Parameterizing Oceanic Turbulence using SmartSim}
\subsection{Motivation}
Climate change poses an ongoing and escalating threat to social and economical welfare, and has been characterized by the United Nations as the defining issue of our time \cite{IPCC_AR5_policymakers}. Beginning with \cite{Manabe1967}, numerical climate models (comprised of physical and biogeochemical models of the atmosphere, ocean, land, and cryosphere) have become an invaluable tool to understand both historical and future change in the Earth's climate. Both numerical models and observational studies confirm that the ocean is the primary sink of the excess heat and carbon dioxide associated with climate change. It is responsible for the delay in realizing the global warming expected from our present CO$_2$ concentrations \cite{Trenberth14} and absorbing between 30 and 40\% of anthropogenic carbon emitted since the dawn of the industrial age \cite{DeVries17, Gruber19}.

The ocean component of these climate models represents a significant computational expense which is primarily a function of the model's spatial resolution. Of the models submitted to the most recent Coupled Model Intercomparison Project \cite{eyring_overview_2016}, most are typically run at spatial resolutions of O(100km) with timesteps of O(1 hour) to accommodate experiments that span centuries to millennia of simulated time. However, these spatial resolutions do not resolve the large-scale, turbulent structures, known as mesoscale eddies, which dominate the ocean's kinetic energy field and transport heat, salt and biogeochemical tracers \cite{JayneMarotzke02, FerrariWunsch09, Chelton11}. Climate-scale ocean general
circulation models (OGCMs) would need to be run at roughly 8 to 16 times higher resolution than present models to resolve mesoscale eddies, increasing the computational cost by a factor of 512 to 4,096. Accurate and robust turbulence \emph{parameterizations}, which estimate the effects of the unresolved eddies, are crucial for providing skillful representations of the ocean \cite{BovilleGent98} and the climate system as a whole.

The partial differential equations (PDEs) solved numerically by OGCMs are a specific application of the Navier-Stokes equations that express the laws of conservation of mass, momentum, and energy in continuum mechanics form. Parameterizations in these equations generally take the form of extra terms that are added to account for physical processes that occur at scales smaller than the model grid. Without additional constraints these terms potentially violate the fundamental conservation laws, or, at minimum, improperly represent the ways in which these quantities are added, removed, or transferred within the model domain.

Expressing the conservation laws in terms of their effects on the kinetic and potential energies of the flow can clarify how these parameterizations should behave. Viewed through this lens, parameterizations should account for how sub-gridscale (``eddy'') energy is exchanged with the energy of the resolved flow. Modern ocean modeling theory thus considers the eddy energy to be a lynchpin variable mediating essentially all of the parameterizations in the model. However, as with all other sub-gridscale variables, a method for obtaining the eddy energy must be developed separately from the solution of the fundamental PDEs. 

A PDE describing the evolution of the eddy kinetic energy can be derived from theory, but includes many terms that cannot be expressed using only quantities associated with a model's resolved flow. The present state-of-the-art method for obtaining the eddy energy thus invokes an extra PDE that approximates its true mathematical form \cite{Jansen2015}. The modifications to the true PDE are severe; many terms are dropped and multiple others are parameterized, limiting the fidelity of the equation to truly represent eddy effects. In this study, we demonstrate an alternative ML-based approach that predicts eddy kinetic energy based on features of the resolved oceanic circulation. 

\subsection{Numerical model description}
This study uses three global configurations of the Modular Ocean Model version 6 (MOM6), an OGCM that has been used for ocean climate simulations (e.g.  \cite{adcroft_gfdl_2019}). The first configuration, referred to as ER, uses a spatial resolution of 1/10$^\circ$ in both the latitudinal and longitudinal directions (about 10 km, resulting in 7.5 million ocean grid points). This resolution is sufficient to resolve mesoscale eddies between the equator and mid-latitudes ($\approx40^\circ$ N/S). The other two configurations use a spatial resolution of 1/4$^\circ$ (about 25 km, resulting in 2.7 million ocean grid points), which falls into the so-called `eddy-permitting' regime where the eddies are partly resolved but turbulence parameterizations are still needed. These two configurations are identical except for the eddy parameterizations that are employed: one uses ``MEKE'' \cite{Jansen2015}, a prognostic eddy kinetic energy equation that represents the current state-of-the-art, while the other uses ``SmartSim-EKE'', the trained neural network (NN) (detailed in Section \ref{subsection:neural_network}) to infer the eddy kinetic energy. The estimates of EKE are then converted to a coefficient used to control the strength of the Gent-McWilliams \cite{gent_isopycnal_1990} parameterization of eddy effects on the resolved circulation.

To generate the data used to train the NN and to provide a benchmark of comparison for EKE, the ER simulation is integrated for 20 years to allow the eddy field to come into equilibrium. EKE at various scales is calculated by coarsening the output by a factor of 2-10. Four 2D diagnostics of the model output are also computed with the coarsened output to train the NN: surface mean kinetic energy (MKE), the Rossby radius of deformation (i.e. an approximate spatial scale of turbulence) normalized by the horizontal grid area, surface relative vorticity (vertical component of the curl of the velocity field), and the column-averaged isopycnal slope (a measure of the potential energy available to generate turbulence).

\subsection{Neural network architecture and training} 
\label{subsection:neural_network}
\begin{figure*}[!htbp]
    \centering
    \includegraphics[width=0.75\textwidth]{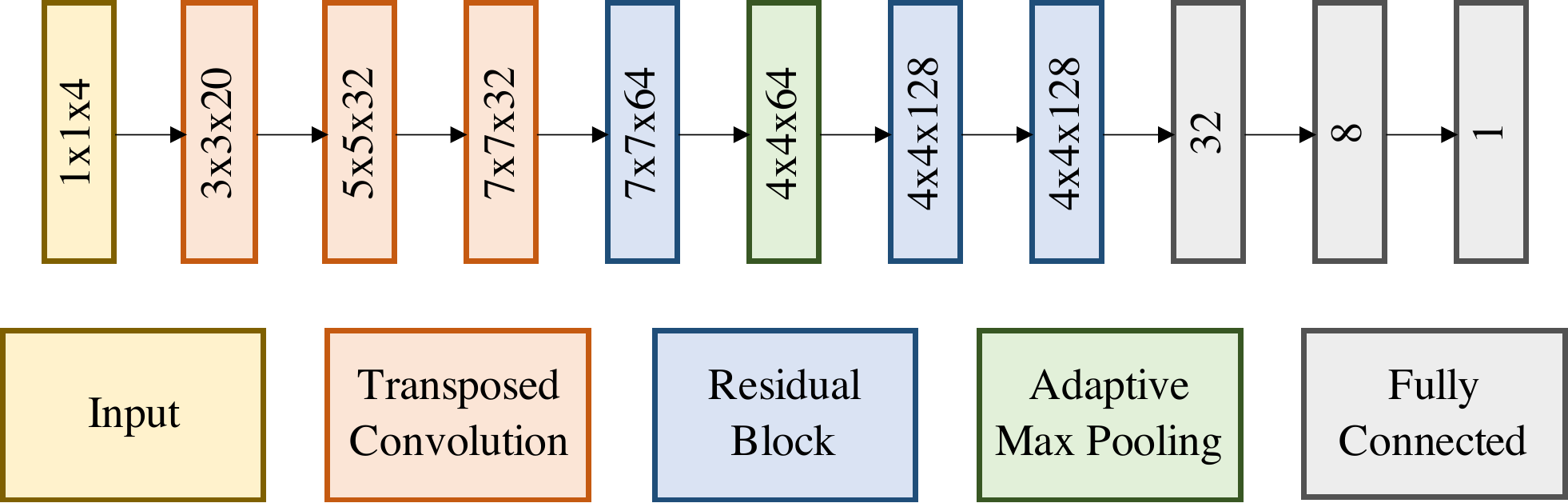}
    \caption{Simplified view of EKEResNet topology. For each constitutive block, the dimension of the output tensor is shown as $W$x$H$x$C$. Residual blocks are implemented as bottleneck blocks. }
    \label{fig:EKEResNet}
\end{figure*}

To predict the EKE value, we used a small NN mainly consisting of residual blocks; these are sequences of 2D convolutional layers, in which skip-connections sum the output of one internal layer to the output of the block: they are typically found in ResNet-derived networks \cite{ResNet, ResNext, WideResNet}. Such building blocks were chosen primarily for two reasons: 2D-convolutions are computed efficiently by the GPUs employed for training and inference, and residual blocks (and their skip-connections) show advantages during the gradient back-propagation phase of the NN training. Since the input of the NN is a single point with 4 features, the first layers are transposed convolutions (also called deconvolutions), which extend the width and height of each sample to 7, so that the convolutional layers in the residual blocks can operate on it. Different types and numbers of residual blocks were tested, and the best results (in terms of time to accuracy and robustness) were obtained with three bottleneck residual blocks \cite{ResNet} performing group convolutions. Attempts to use fewer residual blocks resulted in faster inference, but also in some numeric errors (where EKE could not be computed). The final part of the NN consists of two fully connected layers that output the predicted surface EKE value. Throughout the rest of this work, the NN topology (shown in Figure~\ref{fig:EKEResNet}) will be referred to as EKEResNet.

As mentioned above, each input sample has four features (predictors). The features have different statistical distributions, and an \textit{ad hoc} pre-processing step is needed to avoid numerical problems when training the NN. MKE and isopycnal slope approximately follow a log-normal distribution, requiring a natural log transformation. The distribution of relative vorticity is symmetric, with a very narrow peak around zero, but a range of several orders of magnitude. To reduce the range without losing information around zero, the function
\begin{equation}
\label{eq:preprocessf}
    f_p(x) =
    \begin{cases}
    - \ln \left|x\right| -C &\text{if} x<0, \\
    0 &\text{if}\; x=0, \\
    \ln x +C  &\text{else}
    \end{cases}
\end{equation}
is applied to relative vorticity. Intuitively, the effect of $f_p$ is to apply the logarithm to both the positive and the negative domain. The result obtained on the negative domain, where the absolute value of $x$ has to be taken, is multiplied by $-1$ to ensure monotonicity. This is not sufficient to make the function injective, as values around zero diverge to $\pm \infty$. As injectivity is desirable to avoid loss of information, a constant value $C$ is added (subtracted) to the results obtained on the positive (negative) domain. $C$ is chosen so that $C>\ln{\epsilon}$, where $\epsilon$ is a cutoff parameter representing the smallest non-zero value which can be encountered in the distribution. This value could be enforced by setting $x$ to 0 when $x<\epsilon$, or it could be the machine accuracy corresponding to the floating point precision chosen for the NN training. $C$ was set to 36 for this work.

After the pre-processing step the mean and standard deviation of each feature is stored and then every feature is standardized. Notice that the Rossby radius of deformation is not pre-processed otherwise, as standardizing it is sufficient to avoid numerical problems or loss of accuracy.

\begin{figure}[htbp]
    \centering
    \vspace{-0.5cm}
    \includegraphics[width=\columnwidth]{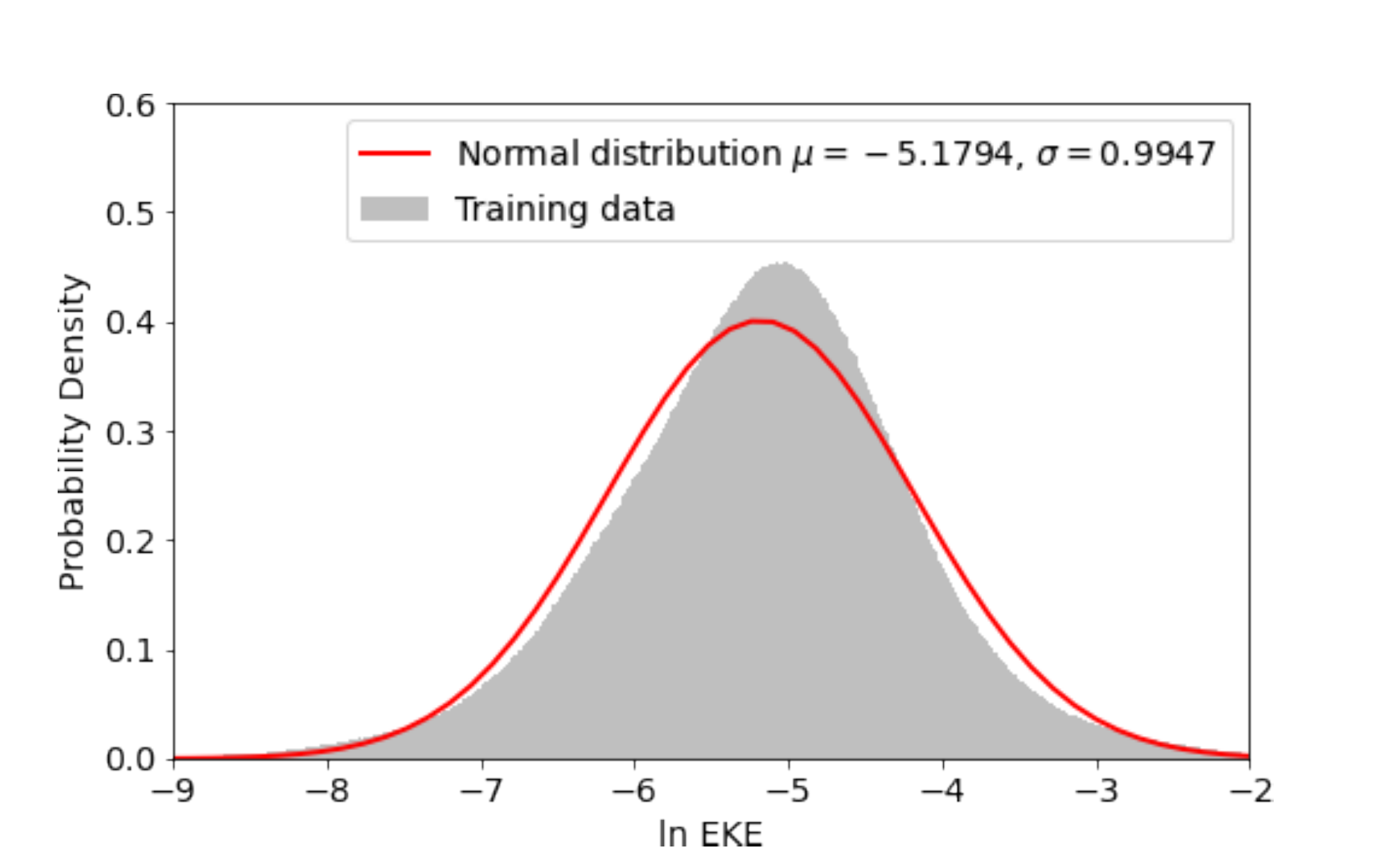}
    \vspace{-0.6cm}
    \caption{Distributions of $\ln(\textnormal{EKE})$ on the whole training dataset and fitted normal distribution.}
    \label{fig:normal_distribution}
\end{figure}

As shown in Figure~\ref{fig:normal_distribution}, the EKE values appear to approximately follow a log-normal distribution, thus the network is trained against $\ln(\textnormal{EKE})$. We store mean and standard deviation of $\ln(\textnormal{EKE})$, as it will be used in a variant of our training strategy.

EKEResNet was trained with Stochastic Gradient Descent, with a learning rate of $4\times10^{-3}$ and a step-wise schedule with peak value reached at the fifth epoch, out of a total of 100. The mean square error was employed as a loss function, with $L2$ penalty of $2 \times 10^{-4}$. The training was performed in parallel on 8 GPUs (Nvidia P100 or V100), with a local batch size of 512 samples. Parallelism was achieved using HPE DL-plugin \cite{mathuriya2018cosmoflow}. 

With the described setting, the NN shows a strong tendency towards overfitting values close to the mean of the distribution. This could be due to the abundance of samples belonging to the $\ln(\textnormal{EKE})$ distribution mean, and the scarcity of samples belonging to the distribution tails. This is a problem not only because the resulting distribution is different from the target one, but also because the most important EKE values for the simulation are those close to the highest values of the distribution, and a model trained with this approach tends to miss them.

\begin{figure}[!htbp]
    \centering
    \vspace{-0.5cm}
    \includegraphics[width=\columnwidth]{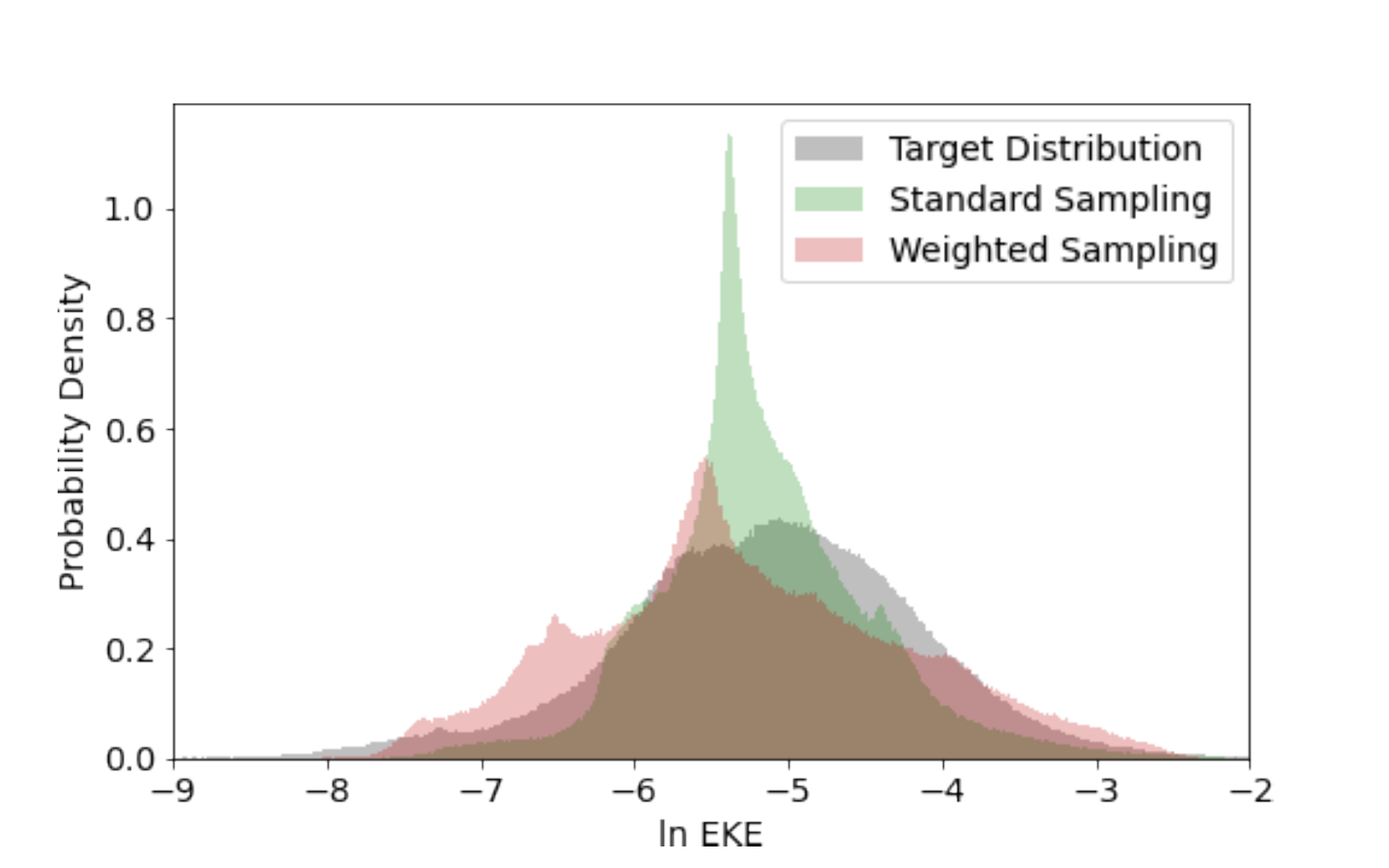}
    \vspace{-0.6cm}
    \caption{Distributions of $\ln(\textnormal{EKE})$ values as predicted by the NN trained with uniform and weighted sampling, compared to ground truth for one test sample.}
    \label{fig:pred_distributions}
\end{figure}

To mitigate this problem, a weighted sampling scheme was used: to each data-point in the training set a weight corresponding to the inverse of its distribution probability density is assigned, this weight is proportional to the probability of drawing each sample during a training epoch. For each epoch, only one-tenth of the data set was used. With this approach, the probability of drawing samples of each portion of the $\ln(\textnormal{EKE})$ domain becomes more uniform, and the distribution of the predicted $\ln(\textnormal{EKE})$ values has a broader peak and extends more to the tails of the distribution, as shown in Figure~\ref{fig:pred_distributions}.

While it is true that the weighted sampling gives a qualitatively better distribution of $\ln(\textnormal{EKE})$, it also has a negative effect on the achieved accuracy: the minimum mean square error attained during the training was $0.55$ using uniform sampling and $0.60$ with weighted sampling.

\subsection{Results}
\label{application-results}

\begin{figure}[ht]
    \centering
    \includegraphics[width=0.9\columnwidth]{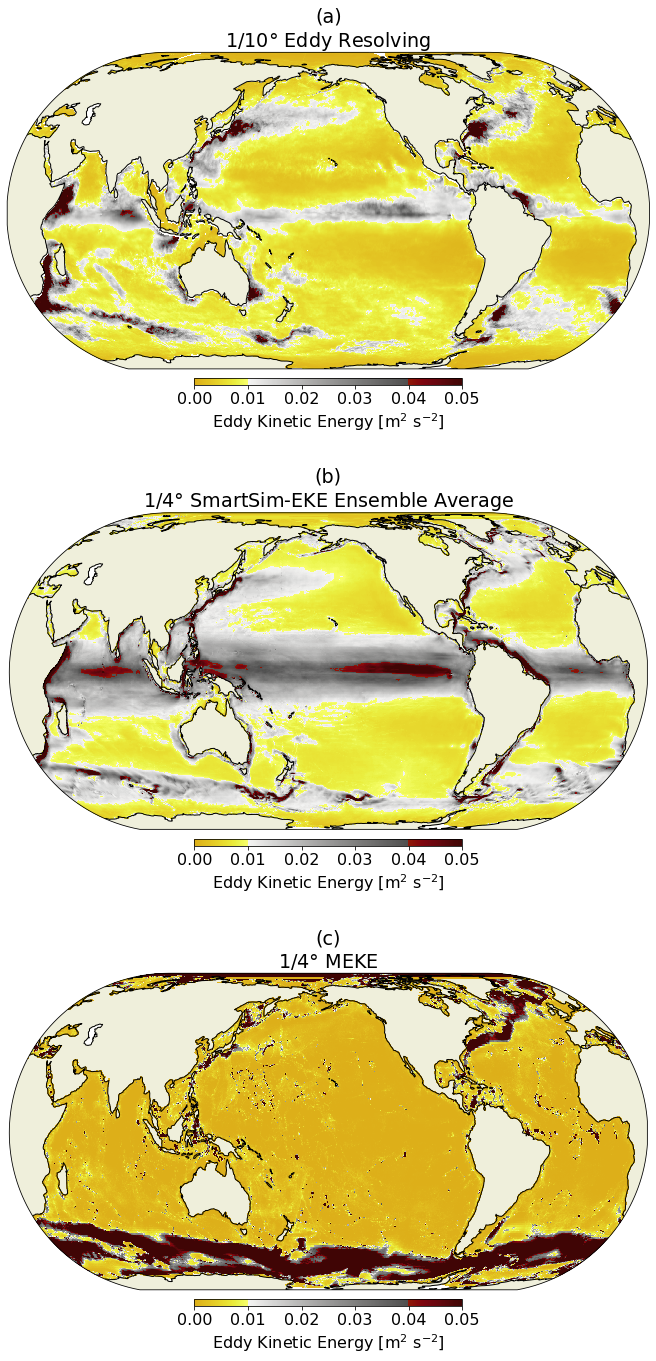}
    \caption{Eddy kinetic energy (EKE), averaged over the last year of each simulation, calculated from the eddy-resolving (ER) 1/10$^\circ$ simulation (a), inferred online using EKEResNet (referred to as SmartSim-EKE), averaged over all 12 ensemble members (b), and the current state-of-the-art MEKE parameterization (c). Both SmartSim-EKE and MEKE use a 1/4$^\circ$ grid which is slightly coarser than the factor of 2 coarsening shown in (a); the ER EKE thus represents a lower-bound on what the `true' EKE should be at that resolution.}
    \label{fig:eke_comparison}
\end{figure}

An ensemble of 12 SmartSim-EKE simulations were run to demonstrate the scalability and performance of SmartSim and to mirror the type of experiments used to characterize uncertainty in weather predictions and climate projections. These ensemble members were branched from a previous 20-year spin-up simulation that used the MEKE parameterization. A spatially-random 0.0001 $^\circ$C perturbation was added to the 3D temperature field of each member to differentiate it from the others in the ensemble. Each member was integrated for 10 additional simulated years using 910 physical cores (10,920 cores total, using a combination of Intel Skylake and Cascadelake processors) with 16 Nvidia P100 GPUs dedicated for the shared Orchestrator (CPU/GPU ratio of about 680).

The online inference portion of MOM6 (i.e. the SmartRedis-based calls to EKEResNet) is called 8 times per simulated day (about every 4 seconds of walltime) on 2.7 million grid points spread across 910 cores. Because MOM6 must wait for every online inference loop to complete, the most accurate way to compare the overall expense of the SmartSim-EKE-based approach is to use the timings of the slowest subdomain. Examining the timings for ensemble member 1 (which is representative of the ensemble as a whole), the total elapsed time in \texttt{put\_tensor} was 24s whereas \texttt{run\_model} was about 2 hours. Based on a total wall time of 136 hours, this represents an overhead of about 6\% compared to the existing MEKE parameterization. We note that this 6\% overhead was also seen when running an ensemble member individually, suggesting that the Orchestrator was not being overtaxed. Integrated over the entirety of the ensemble simulation, approximately 970 billion inferences were performed resulting in an average rate of 1.86 million online inferences per second.

SmartSim-EKE and MEKE are compared to the ``true" EKE calculated directly from ER (Figure \ref{fig:eke_comparison}). The ER simulation correctly shows elevated EKE in the regions where vigorous eddy activity is expected: western boundary currents (e.g. the Gulf Stream and the Kuroshio currents), the Southern Ocean, and the eastern equatorial Pacific. In comparison, SmartSim-EKE generally overestimates the extent of the equatorial EKE, but otherwise reasonably captures the magnitude and large-scale patterns seen in ER. Despite these differences, SmartSim-EKE is a clear improvement over MEKE, which grossly overestimates EKE in the Southern Ocean and the Gulf Stream and generates unrealistically low values of EKE nearly everywhere else. MEKE also crucially overestimates EKE in the Arctic Ocean, which is one of the more quiescent areas of the ocean. This suggests that some of the omitted or parameterized terms in MEKE's prognostic EKE equation result in significant structural biases.

\begin{figure}[ht!]
    \centering
    \includegraphics[width=0.9\columnwidth]{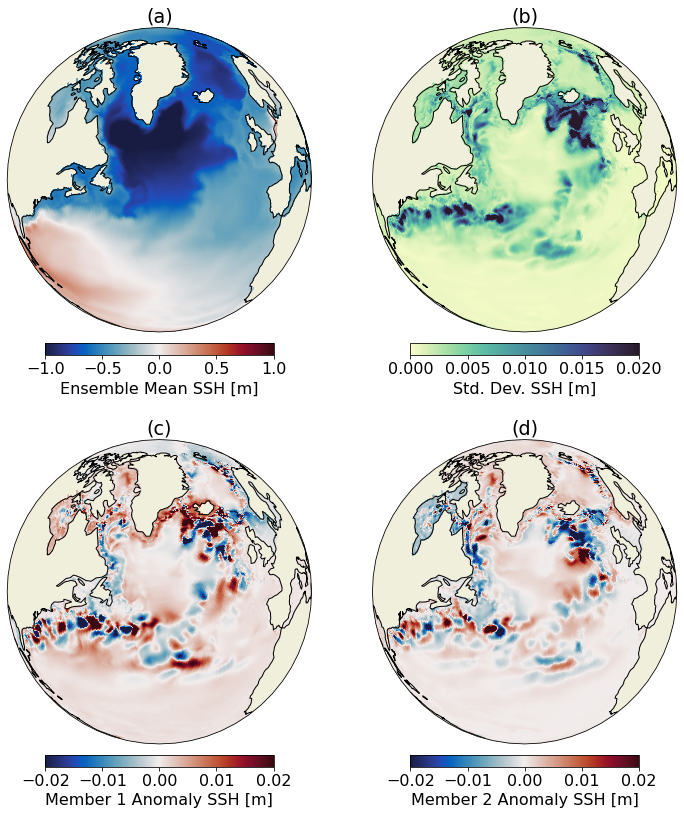}
    \caption{Mean (a) and variability (b) of sea surface height (roughly the streamlines of the large-scale flow) across the SmartSim-EKE ensemble as diagnosed by sea surface height on the last day of the simulation. Panels (c) and (d) show the difference in SSH between ensemble members 1 (c) and 2 (d) and the ensemble mean. } 
    \label{fig:ensemble_ssh}
\end{figure}

The 10-year integrations shown here are not sufficient to evaluate whether the improved representations of EKE result in a more skillful representation of the ocean for weather prediction or climate. However, they are sufficient to demonstrate that the inclusion of machine-learning still leads to stable, realistic simulations of the large-scale ocean circulation. In the Atlantic basin, the Gulf Stream separates the clockwise flow of the subtropical gyre from the subpolar North Atlantic gyre. These two gyres can be seen clearly in the ensemble average of SSH (Figure \ref{fig:ensemble_ssh}a) as positive (red) and negative (blue) colors, respectively. The standard deviation of SSH (Figure \ref{fig:ensemble_ssh}) is also elevated (blue) in the expected locations where strong eddy activity modulates the path of the North Atlantic Current. The SSH anomalies in two of the ensemble members (Figures \ref{fig:ensemble_ssh}c,d) show trains of positive and negative anomalies indicating the presence of eddies. This suggests that the SmartSim-EKE parameterization is not so strong that it is suppressing resolved eddies, a well-known pitfall in eddy-permitting simulations \cite{Hallberg13}. We note that all 12 ensemble members completed their 10-year simulations with no evidence of numerical instability.

The MOM6 SmartSim-EKE ensemble demonstrated in this study serves as an important step towards longer, climate-scale simulations, including in free-running, fully coupled (ice, ocean, atmosphere, and land) configurations. Such centennial-length experiments are currently being designed and will be evaluated to understand whether SmartSim-EKE improves the representation of the climate when using a 1/4$^\circ$ ocean component.

\section{Discussion}\label{section:Discussion}
This study introduces a new software solution, SmartSim, that is composed of an Infrastructure Library and a Client Library that can couple existing High Performance Computing (HPC) simulations written in Fortran/C/C++ to Machine Learning (ML) and data analysis libraries online and at scale. We demonstrated that, with minimal code changes, SmartSim clients can leverage the SmartRedis API to support the remote execution of TensorFlow, Keras, ONNX, and PyTorch models and scripts for distributed, online inference.

In addition to the describing the SmartSim architecture, we demonstrated the particular use case of integrating SmartSim with Modular Ocean Model 6 (MOM6), which is written in Fortran. We replaced an existing parameterization for eddy kinetic energy with a data driven, ML model. The JIT-traced, PyTorch model was queried at runtime (online) for the prediction of eddy kinetic energy by the Fortran SmartRedis Client. Additionally, we showed that SmartSim is capable of running ensembles of global ocean simulations utilizing the same ML infrastructure, with minimal impact on ensemble member (simulation) runtime. As stated, we believe the results shown here are the first demonstrations of using ML within freely running, realistic, global simulations of the ocean. 

SmartSim facilitates the convergence of AI and numerical simulation workloads by removing the necessity for file I/O, providing clients to seamlessly connect applications across programming languages, and utilizing a ML library agnostic API between workloads. As mentioned, previous efforts to utilize ML in simulation models have not addressed impediments to the actual utilization of ML. Approaches that re-create ML libraries in simulation languages, embed large ML libraries or python interpreters, or use the filesystem as an intermediary lack the flexibility to support and benefit from the rapid advancements in the data science ecosystem.

Due to the massive size of compute resources, HPC applications have historically maintained highly controlled strategies for communication and data access. The result has been optimized, yet constrained, data flow driven by MPI communication barriers. Data extraction has almost entirely relied on parallel, networked filesystems which introduce meaningful delays in simulation model development, analysis and enhancement. Users have not, without major architectural changes, been able to easily couple applications across languages, runtimes, and heterogeneous processor types (CPU/GPU). The loosely coupled nature of the data communication in SmartSim enables new paradigms in application coupling, computational steering, real-time analysis, and the utilization of ML at scale.

The authors recognize that loosening the constraints on data communication in HPC applications may lead to degradation in application performance at extreme scales. We believe the added flexibility of communication is of greater importance given the types of applications and systems that are enabled through this approach. In addition, we show SmartSim scales linearly to thousands of processors in multiple configurations on modern, heterogeneous HPC systems. Our scaling study, application, and SmartSim codebase are all available with instructions for reproduction of this work.

In future work, the authors plan to investigate the utilization of SmartSim for continuous online training of ML models as well as support new features such as flash storage and asynchronous data communication. In conjunction, integrations of SmartRedis data-structures into popular open source data formats (ex. Xarray, VTK) and libraries (ex. Ray, Dask) will be explored. 

\section*{Acknowledgment}
The authors would like to acknowledge Elliot Ronaghan for his help with application performance, and RedisLabs for their continued support.
% \section*{References}

% \begin{thebibliography}
\bibliographystyle{IEEEtran}
\bibliography{shao,partee,ellis,rigazzi}

% Appendix
\newpage
\newpage

\end{document}